\documentclass[letterpaper, 11pt, onecolumn, titlepage, floatfix, fleqn,
  preprintnumbers, showpacs, showkeys]{revtex4-1}

\usepackage[colorlinks=true, linkcolor=blue, citecolor=blue,
            filecolor=blue, urlcolor=blue]{hyperref}
\usepackage{graphicx,moreverb}

\newcommand{\q}[2]{\ensuremath{#1\ \mathrm{#2}}}
\newcommand{\be}{\begin{equation}}
\newcommand{\ee}{\end{equation}}
\newcommand{\bea}{\begin{eqnarray}}
\newcommand{\eea}{\end{eqnarray}}
\newcommand{\dt}{\ensuremath{\Delta t}}

\newcommand{\PN}[1]{\ensuremath{P\left(\frac{#1}{\sigma}\right)}}
\newcommand{\e}[1]{\ensuremath{\exp{\left[-\frac{1}{2}\left(\frac{#1}{\sigma}\right)^2\right]}}}

\usepackage{Sweave}
\begin{document}

\date{\today}

\title{Diffusion model for the time evolution of particle loss rates
  in collimator scans: a method for measuring stochastic transverse
  beam dynamics\\ in circular accelerators}

\author{Giulio~Stancari}
\affiliation{Fermi National Accelerator Laboratory, P.O. Box 500,
  Batavia, Illinois 60510, USA}

\begin{abstract}
  A diffusion model of the time evolution of loss rates caused by a
  step in collimator position is presented. It builds upon the model
  of Ref.~\cite{Seidel:1994} and its assumptions: (1)~constant
  diffusion rate within the range of the step and (2)~linear halo
  tails. These hypotheses allow one to obtain analytical expressions
  for the solutions of the diffusion equation and for the
  corresponding loss rates vs.\ time. The present model addresses some
  of the limitiations of the previous model and expands it in the
  following ways: (a)~losses before, during, and after the step are
  predicted; (b)~different steady-state rates before and after are
  explained; (c)~determination of the model parameters (diffusion
  coefficient, tail population, detector calibration, and background
  rate) is more robust and precise. These calculations are the basis
  for the measurement of transverse beam diffusion rates as a function
  of particle amplitude with collimator scans. The results of these
  measurements in the Tevatron will be presented in a separate report.
\end{abstract}

\preprint{\href{http://inspirebeta.net/search?p=find+r+fermilab-fn-0926-apc}
               {FERMILAB-FN-0926-APC}}

\maketitle

\tableofcontents
\clearpage


\section{Introduction}

\setkeys{Gin}{width=6in}
\begin{figure}[b]
\includegraphics{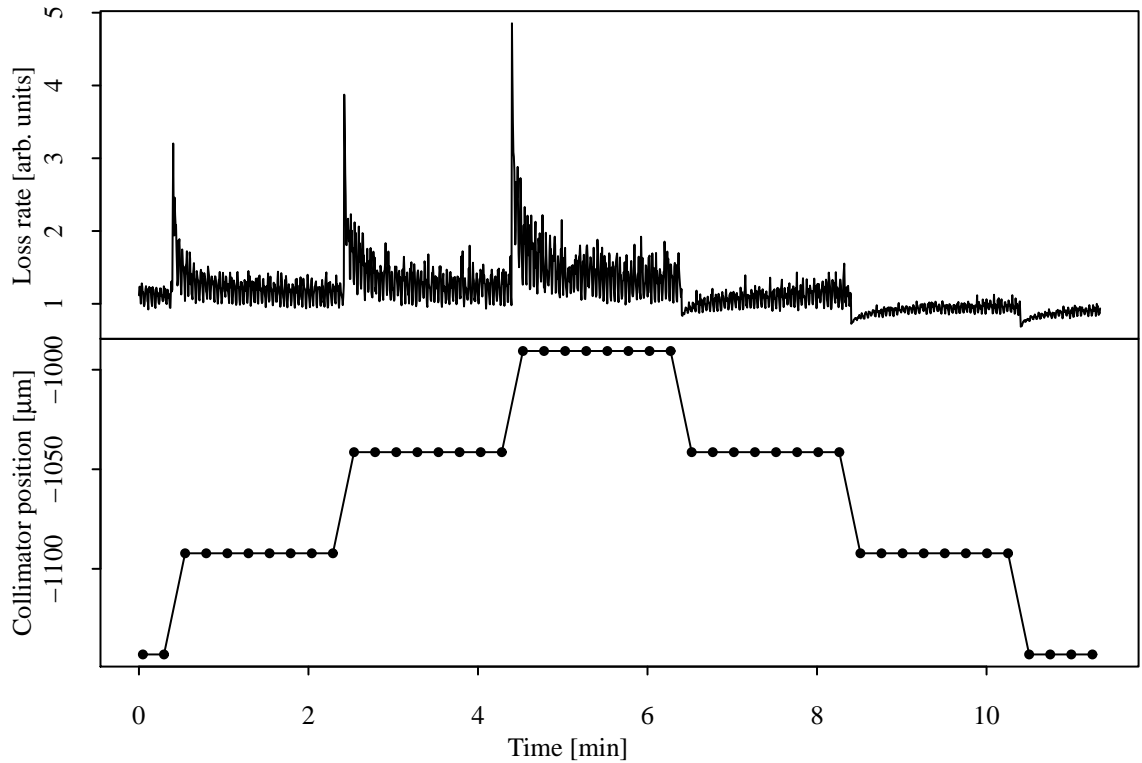}
\caption{Example of loss rate data taken during a collimator scan in
  Tevatron Store~8749 (20 May 2011): local loss rate (top, device
  \texttt{T:LF480}); collimator position (bottom, device
  \texttt{T:F48VCP}). The collimator steps take about 0.2~s. In
  this example the collimator position was recorded only every 15~s.}
\label{fig:data.example}
\end{figure}

Phenomena related to the stochastic transverse beam dynamics in
circular accelerators can be described in terms of particle
diffusion~\cite{Lichtenberg:1992, Chen:PRL:1992,
  Gerasimov:FERMILAB:1992, Zimmermann:PA:1994, Sen:PRL:1996}. It was
demonstrated that these effects can be observed with collimator
scans~\cite{Seidel:1994}. Usually, collimator jaws are the devices
that are closest to the beam and they define the machine aperture. If
they are moved towards the beam center in small steps, typical spikes
in the local shower rate are observed, which approach a new
steady-state level with a characteristic relaxation time. When
collimators are retracted, on the other hand, a dip in losses is
observed, which also tends to a new equilibrium level
(Figure~\ref{fig:data.example}). A detailed description of the
Tevatron collimation system can be found in
Ref.~\cite{Mokhov:JINST:2011}.

These phenomena have been used to estimate the diffusion rate in the
beam halo in the SPS at CERN~\cite{Burnod:CERN:1990}, in HERA at
DESY~\cite{Seidel:1994}, and in RHIC at BNL~\cite{Fliller:PAC:2003}.
Similar measurements were carried out at the Tevatron in~2011. Besides
the interest in characterizing the beam dynamics of colliding beams,
these measurements were motivated by the study of the effects of the
novel hollow electron beam collimator~\cite{Stancari:PRL:2011}.

Here we present a more complete model of beam evolution under
diffusion. It will serve as the basis for interpreting of Tevatron
data. Previous models are extended to explain the behavior of losses
before, during, and after the collimator step. This allows one to
extract the diffusion rate in a more robust way, by taking into
account not only the relaxation time, but also the steady-state loss
rates before and after the step and the peak or dip value. The
analysis of Tevatron data will be presented in a separate report. This
model can also be applied to the dynamics of beams in the LHC.

\section{Model}

Following Ref.~\cite{Seidel:1994}, we consider the evolution in
time~$t$ of a beam of particles with phase-space density~$f(J,t)$
described by the diffusion equation:
\be
\partial_t f = \partial_J \left( D \, \partial_J f \right)
\ee
where~$J$ is the Hamiltonian action and~$D(J)$ the diffusion
coefficient. The particle flux at a given location $J=J'$ is $\phi =
-D \cdot \left[ \partial_J f \right]_{J=J'}$.

During a collimator step, the action~$J_c = x^2_c / \beta_c$,
corresponding to the collimator position~$x_c$ at a ring location
where the amplitude function is~$\beta_c$, changes from its initial
value~$J_{ci}$ to its final value~$J_{cf}$ during a time~\dt. The step
in action is $\Delta J \equiv J_{cf} - J_{ci}$. In the
Tevatron, typical steps are \q{50}{\mu m} in 0.2~s, and the amplitude
function is tens of meters. The behavior of~$J_c(t)$ can be modeled,
for instance, by a linear function connecting~$J_{ci}$ with~$J_{cf}$:
\be
J_c(t) = \left\{
\begin{array}{ll}
J_{ci} & t \leq 0 \\
J_{ci} + (J_{cf}-J_{ci}) \cdot t / \dt & 0 < t < \dt \\
J_{cf} & \dt \leq t
\end{array}
\right.
\ee

It is assumed that the collimator steps are small enough so that the
diffusion coefficient can be treated as a constant in that
region. This hypothesis is justified by the fact that the fractional
change in action is of the order of $\Delta J_c / J_c \sim (2)
(\q{25}{\mu m}) / (\q{2}{mm}) = 2.5\%$. Because the diffusion
coefficient is a strong function of action ($D \sim J^4$), this
translates into a variation of 10\% in the diffusion rate, an
acceptable systematic in a quantity that varies by orders of
magnitude. If~$D$ is constant, the diffusion equation becomes
\be
\partial_t f = D \, \partial_{JJ} f .
\ee

With these definitions, the particle loss rate at the collimator is
\be
L = -D \cdot \left[ \partial_J f \right]_{J=Jc}.
\label{eq:loss.rate}
\ee
Particle showers caused by the loss of beam are measured with
scintillator counters placed close to the collimator jaw. The observed
shower rate is parameterized as follows
\be
S = kL + B,
\label{eq:shower.rate}
\ee
where~$k$ is a normalization constant including detector acceptance
and efficiency and~$B$ is a background term which includes, for
instance, the effect of residual activation. Both~$k$ and~$B$ are
assumed to be independent of collimator position and time during the
scan.

\section{Boundary conditions}

\setkeys{Gin}{width=3.25in}
\begin{figure}[b]
\includegraphics{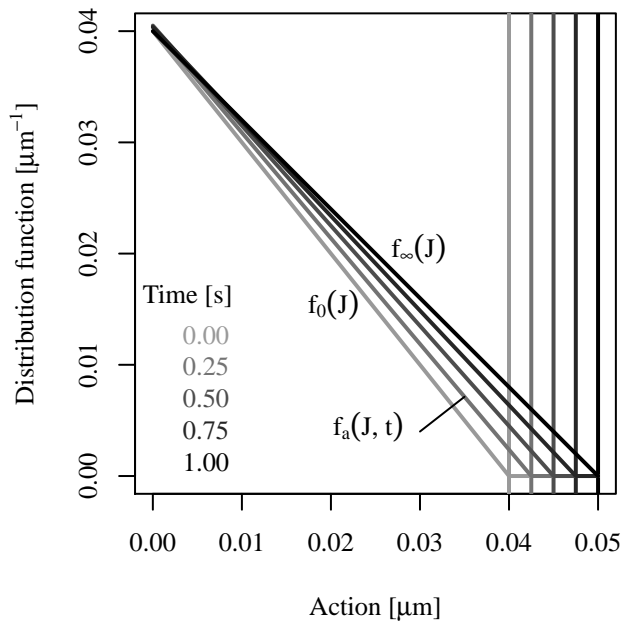}
\caption{Illustration of boundary conditions in the case of an
  outward collimator step: initial distribution~$f_0(J)$,
  intermediate asymptotic distributions~$f_a(J,t)$, final
  distribution~$f_\infty(J)$. The vertical lines represent the
  positions of the collimator vs.\ time. Parameters in this example
  are $J_{ci} = \q{0.04}{\mu m}$, $J_{cf} = \q{0.05}{\mu
    m}$, $A_i = \q{1}{\mu m^{-2}}$, $A_f = \q{0.8}{\mu
    m^{-2}}$, $\Delta t = \q{1}{s}$.}
\label{fig:boundaries}
\end{figure}

The collimator is treated as a perfect absorber, so that the boundary
condition for the phase-space density becomes
\be f(J, t) = 0 \ \ \
\mbox{for $J \geq J_c$}.
\label{eq:boundary}
\ee
We assume that cancellation of the particle flux at $J=0$ is
automatically satisfied by $D(0) \simeq 0$, so no boundary condition
is imposed there. This greatly simplifies the form of Green's function
(see below).

As initial conditions and asymptotic behavior for the phase-space
density we use linear functions of action:
\be
f_0(J) \equiv f(J,0) = \left\{
\begin{array}{ll}
(J_{ci}-J)\cdot A_i & J < J_{ci} \\
0 & J_{ci} \leq J
\end{array}
\right.
\label{eq:asym0}
\ee
\be
f_\infty(J) = f(J,\infty) = \left\{
\begin{array}{ll}
(J_{cf}-J)\cdot A_f & J < J_{cf} \\
0 & J_{cf} \leq J
\end{array}
\right. ,
\label{eq:asym1}
\ee
where~$A_i$ and~$A_f$ are constants. This is the essential hypothesis
that allows one to obtain analytical solutions for the time evolution
of the distribution function. It is justified by considering these
expressions as the first term in the Taylor expansion of the beam
tails. A linear behavior of the asymptotic solution also gives a
constant steady-state flux.

In this respect, the present model differs from that of
Ref.~\cite{Seidel:1994}. We allow the slopes of the initial and final
distributions to be different. This is necessary to explain the
difference in the steady-state loss rates~$L$ before and after the
collimator step. If one assumes the same diffusion coefficient and the
same slope before and after, then one can only predict the same
steady-state rate. Including the measured steady-state loss rates
before and after the collimator step helps to disentangle the effects
of population and diffusion, and to give a physical meaning to the
model parameters.

Because of its linearity, a solution of the diffusion equation can be
found using the method of Green's functions:
\be
f(J, t) = f_a +
  \int_{0}^{J_c} \left( f_0 - f_a \right) \cdot G(J, J', t) \, dJ',
\label{eq:solution}
\ee
where $G(J, J', t)$ is Green's function for the given problem, $f_0$
(Eq.~\ref{eq:asym0}) is the initial distribution, and $f_a$ is an
asymptotic solution. We are looking for a model of losses not only
before and after the collimator step, but also as the collimator is
moving. In this respect, too, this model extends that of
Ref.~\cite{Seidel:1994}. For this reason, we use $f_a = f_\infty$
(Eq.~\ref{eq:asym1}) for $J_c = J_{cf}$ (i.e., $t \geq \Delta t$) and
\be
f_a(J,t) = \left\{
\begin{array}{ll}
(J_c-J)\cdot A_c & J < J_c \\
0 & J_c \leq J
\end{array}
\right. ,
\label{eq:asym2}
\ee
as the collimator moves ($0 < t < \Delta t$). The parameter $A_c(t)$
is chosen to vary linearly between $A_i$ and~$A_f$,
\be
A_c(t) = \left\{
\begin{array}{ll}
A_i & t \leq 0 \\
A_i + (A_c-A_i) \cdot t / \dt & 0 < t < \dt \\
A_f & \dt \leq t
\end{array}
\right. ,
\ee
so that the asymptotic solution transitions smoothly from
Eq.~\ref{eq:asym0} to Eq.~\ref{eq:asym1}.  The initial and asymptotic
solutions are illustrated in Figure~\ref{fig:boundaries}.

The basic kernel for the diffusion equation is
\be
K(J,J',t) = \frac{1}{\sqrt{2\pi} \sigma} \e{J-J'},
\ee
with $\sigma \equiv \sqrt{2Dt}$. To satisfy the boudary condition at
the collimator (Eq.~\ref{eq:boundary}), an antisymmetric Green's
function can be used:
\bea
G(J, J', t) & = & \left\{ \e{(J_c-J')-(J_c-J)} - \e{(J_c-J')+(J_c-J)}
\right\}, \nonumber \\
& & \cdot \frac{1}{\sqrt{2\pi} \sigma}
\eea
so that
\be
G(J=J_c, J', t) = G(J, J'=J_c, t) = 0.
\ee
The requirement that the solution be zero beyond the collimator position,
\be
G(J, J', t) = 0 \ \ \ \mbox{if $J_c < J$ or $J_c < J'$},
\ee
is automatically satisfied by limiting the integration region
between~$0$ and~$J_c$ (Eq.~\ref{eq:solution}). Imposing additional
boundary conditions at $J=0$ would require~$G$ to be an infinite
series. The analytical approximation used here does not constrain the
phase-space density or its gradient at the origin, but it turns out
\emph{a posteriori} that~$f(0,t)$ does not vary significantly if
$f_0(0) \simeq f_\infty(0)$, which is what one would expect for a
collimator step affecting the beam halo and not the beam core. Green's
function also satisfies the general symmetry property $G(J, J', t) =
G(J', J, t)$. We also note its asymptotic behavior in time: $G(J, J',
0) = \delta(J-J')$ and $G(J, J', \infty) = 0$, which justifies the
physical interpretation of~$f_0$ and~$f_a$ in Eq.~\ref{eq:solution} as
initial and asymtotic solutions.

\section{Solutions}

\setkeys{Gin}{width=6in}
\begin{figure}[b]
\includegraphics{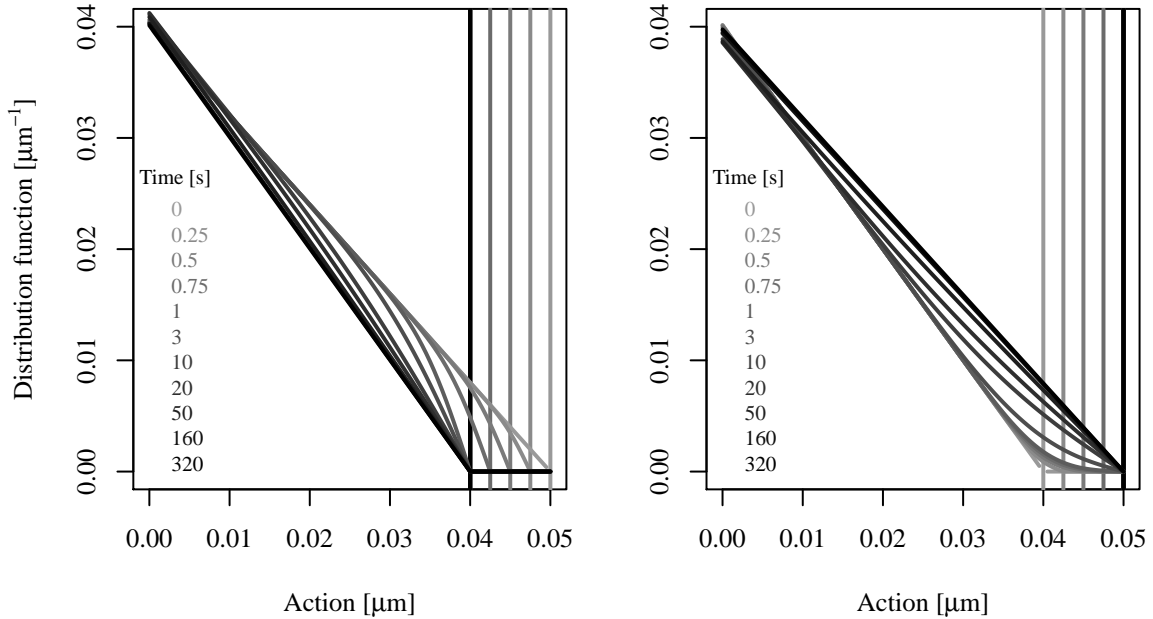}
\caption{Evolution of distribution function during collimator step:
  $f_I(J,t)$ (inward, left) and $f_O(J,t)$ (outward, right). The
  vertical lines represent the positions of the collimator vs.\
  time. Collimator action varies between $J_{ci} = \q{0.05}{\mu
    m}$ and $J_{cf} = \q{0.04}{\mu m}$ in the inward case
  (viceversa in the outward case) in a time $\Delta t =
  \q{1}{s}$. The slopes of the tails are $A_i = \q{0.8}{\mu
    m^{-2}}$ and $A_f = \q{1}{\mu m^{-2}}$ in the inward case
  (viceversa outwards). The diffusion coefficient is $D =
  \q{10^{-5}}{\mu m^2/s}$.}
\label{fig:distrib}
\end{figure}

By setting up the diffusion model in the way described above,
solutions can be expressed analytically through
Eq.~\ref{eq:solution}. It is convenient to treat the cases of inward
($J_{cf} < J_{ci}$) and outward ($J_{ci} < J_{cf}$) movement
separately. In the inward case, the integrand is $f_0 - f_a =
A_i(J_{ci} - J) - A_c(J_c-J)$. In the outward case, it is convenient
to divide the integral into two parts:
\be
\int_0^{J_c} = \int_0^{J_{ci}} + \int_{J_{ci}}^{J_c} .
\ee
This is done because $f_0$ is null beyond the initial collimator
position: $f_0 - f_a = -f_a$ (see also
Figure~\ref{fig:boundaries}). To express the primitive of the Gaussian
function, we use the cumulative Gaussian distribution function~$P(x)$,
defined in Appendix~\ref{sec:formulas}. (Another possible choice is
the so-called error function.) Integration yields the solutions of the
diffusion equation in the two cases, $f_I(J,t)$ (inward step) and
$f_O(J,t)$ (outward), subject to the boundary conditions specified
above:
\bea
f_I(J,t) = & - & A_i (J_{ci} + J - 2J_c) \\
 & + & 2 \PN{J_c-J} A_i (J_{ci} - J_c)
 \nonumber \\
 & - & \PN{-J} \left[A_i(J_{ci}-J) -
   A_c(J_c-J)\right] \nonumber \\
 & + & \PN{J-2J_c} \left[A_i(J_{ci}+J-2J_c) +
   A_c(J_c-J)\right] \nonumber \\
 & + & \frac{\sigma}{\sqrt{2\pi}} (A_c-A_i) \left\{ \e{J} - \e{J-2J_c}
 \right\} \nonumber
\eea
\bea
f_O(J,t) = & + & \PN{J_{ci}-J} A_i (J_{ci} - J) \\
 & - & 2 \PN{J_{ci}+J-2J_c} A_i (J_{ci} + J - 2J_c)
 \nonumber \\
 & - & \PN{-J} \left[A_i(J_{ci}-J) -
   A_c(J_c-J)\right] \nonumber \\
 & + & \PN{J-2J_c} \left[A_i(J_{ci}+J-2J_c) +
   A_c(J_c-J)\right] \nonumber \\
 & + & \frac{\sigma}{\sqrt{2\pi}} A_i \left\{ \e{J_{ci}-J} - \e{J_{ci}+J-2J_c}
 \right\} \nonumber \\
 & + & \frac{\sigma}{\sqrt{2\pi}} (A_c-A_i) \left\{ \e{J} - \e{J-2J_c}
 \right\} \nonumber
\eea
Some examples of the evolution of the phase-space density described by
these functions are shown in Figure~\ref{fig:distrib}.  A few
representative snapshots in time are chosen: during collimator
movement; a short time after the step, with a time scale determined by
$t_s = \left|J_{ci}-J_{cf}\right|^2/D = \q{10}{s}$;
and a long time after the step, with a characteristic time $t_l =
\left[\min{(J_{ci},J_{cf})}\right]^2 / D =
\q{160}{s}$.

\section{Time evolution of losses}

\setkeys{Gin}{width=6in}
\begin{figure}[b]
\includegraphics{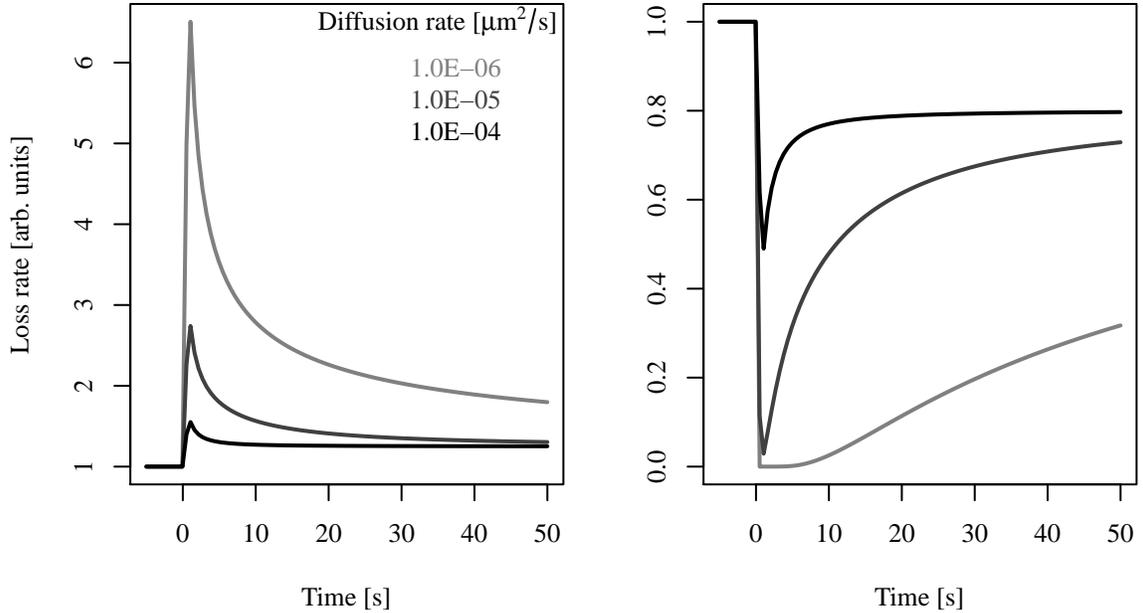}
\caption{Calculated evolution of loss rates~$L(t)$ during a collimator
  step according to Eqs.~\ref{eq:loss.rate}, \ref{eq:gradI}
  and~\ref{eq:gradO}: inward (left) and outward (right). Collimator
  action varies between $J_{ci} = \q{0.05}{\mu m}$ and $J_{cf} =
  \q{0.04}{\mu m}$ in the inward case (viceversa in the outward
  case) in a time $\Delta t = \q{1}{s}$ (see also
  Figure~\ref{fig:distrib}). The effect of 3 different
  values of the diffusion coefficient~$D$ is shown.  The slopes of the
  tails are scaled so that the initial and final steady-state loss
  rates are the same in all cases: $A_i = 1/D$, $A_f = A_i J_{ci} /
  J_{cf}$.}
\label{fig:losses}
\end{figure}

Local losses are proportional to the gradient of the distribution
function at the collimator (Eq.~\ref{eq:loss.rate}). The partial
derivatives of the phase-space density with respect to action are the
following:
\bea
\partial_J f_I(J,t) = & - & A_i
 + (A_i-A_c) \left[ \PN{-J} + \PN{J-2J_c} \right] \\
 & + & \frac{1}{\sqrt{2\pi} \sigma} \left\{ 2 A_i (J_c - J_{ci})
   \e{J_c-J} + \right. \nonumber \\
& & \left.  (A_i J_{ci} - A_c J_c) \left[ \e{J} + \e{J-2J_c}
   \right] \right\}, \nonumber
\eea
\bea
\partial_J f_O(J,t) = & - & A_i \left[ \PN{J_{ci}-J} +
  \PN{J_{ci}+J-2J_c} \right] \\
& + & (A_i-A_c) \left[ \PN{-J} + \PN{J-2J_c} \right] \nonumber \\
& + & \frac{A_i J_{ci} - A_c J_c}{\sqrt{2\pi} \sigma} \left\{ \e{J} +
  \e{J-2J_c} \right\}. \nonumber
\eea

The value of the gradient at the collimator is therefore
\bea
\label{eq:gradI}
\partial_J f_I(J_c, t) = & - & A_i + 2(A_i - A_c) \PN{-J_c} \\
 & + & \frac{1}{\sqrt{2\pi} \sigma} \left\{ -2A_i(J_{ci}-J_c) + 2 (A_i
J_{ci}-A_c J_c) \e{J_c} \right\}, \nonumber
\eea
\bea
\label{eq:gradO}
\partial_J f_O(J_c, t) = & - & 2 A_i \PN{J_{ci}-J_c}
 + 2 (A_i - A_c) \PN{-J_c} \\
 & + & 2 \frac{A_i J_{ci} - A_c J_c}{\sqrt{2\pi} \sigma} \e{J_c}. \nonumber
\eea

These are the functions that are used to model the measured shower
rates (Eqs.~\ref{eq:loss.rate} and~\ref{eq:shower.rate}). As expected,
both functions tend to $-A_i$ for $t\to 0$ and to $-A_f$ as $t \to
\infty$. For the $t \to 0$ limit to hold for the outward solution, it
is necessary that $J_c \to J_{ci}$ faster than $\sqrt{t}$, which is
satisfied by the linear approximation adopted here; otherwise, the
slope will tend to zero.

These functions explain the data very well. In the transient region,
after the collimator has reached its final position, they agree with
those calculated in Ref.~\cite{Seidel:1994}. Their main feature is a
decay proportional to the square root of time (through the
parameter~$\sigma$), as is typical of diffusion processes. A few
examples are plotted in Figure~\ref{fig:losses}.

\section{Comments on parameter estimation}

Having a model that describes the data before, during, and after the
collimator step has several advantages. The products~$k D A_i + B$ and
$k D A_f +B$ are determined by the steady-state loss rates. If a data
set includes measurements of several steps at different amplitudes,
the parameters~$k$ and~$B$, which are independent of~$J$ and~$t$, can
be determined separately. Therefore, steady-state rates constrain the
products~$D A_i$ and~$D A_f$ at each step.

The value of the diffusion coefficient~$D$ is constrained both by the
peak (or dip) value relative to the steady-state rate and by the
duration of the transient through the parameter~$\sigma$. In fact, the
peak (or dip) value of the loss rate is achieved when the collimator
reaches its final position ($J_c = J_{cf}$, $t = \Delta t$). At this
point, neglecting the background, the loss rate is
\be
S(\Delta t) \simeq k D A_i \left[ 1 \pm \frac{\left| \Delta J
    \right|}{\sqrt{\pi \, D \, \Delta t}} \right] ,
\ee
whereas $S(0) \simeq k D A_i$. It follows that an estimate of the
diffusion rate is
\be
D \simeq \frac{(\Delta J)^2}
              {\pi \, \Delta t \, \left[ S(\Delta t)/S(0) - 1 \right]^2} .
\ee
On the other hand, losses relax with a typical time constant that
depends on the diffusion rate and on the magnitude of the step. One
may define the characteristic time~$t_\pi$ so that
\be
\frac{\left| \Delta J \right|}{\sqrt{\pi \, D \, t_\pi}} \simeq
\frac{1}{\sqrt{\pi}} = 0.56,
\ee
meaning that the magnitude of the transient at time~$t_\pi$ is about
half of that of the steady-state rate. Therefore, we have an
independent estimate of~$D$:
\be
D \simeq \frac{(\Delta J)^2}{t_\pi}.
\ee
If only the $t > \Delta t$ data is considered, as is done in
Ref.~\cite{Seidel:1994}, this is the only available information
on~$D$. In addition, in this case, the diffusion coefficient is highly
correlated with the steady-state parameters.

These rough estimates of the model parameters can be used as initial
guesses in a least-squares fit of the data.

\appendix

\section{Useful formulas}
\label{sec:formulas}

To express the solutions of the diffusion equation, we use the
cumulative Gaussian distribution function~$P(x)$, defined as follows:
\be
P(x) \equiv \frac{1}{\sqrt{2\pi}} \int_{-\infty}^x \exp{(-z^2/2)} \,
dz .
\ee
Therefore, $P(-\infty)=0$, $P(0)=1/2$, and $P(\infty)=1$. By
definition, the integral of a Gaussian function can be expressed as
follows:
\be
\frac{1}{\sqrt{2\pi}\sigma} \int_{z_1}^{z_2} \e{z-z_0} \, dz =
P\left( \frac{z_2-z_0}{\sigma} \right) -
P\left( \frac{z_1-z_0}{\sigma} \right) .
\ee
For our purposes, another useful integral is
\bea
\frac{1}{\sqrt{2\pi}\sigma} \int_{z_1}^{z_2} z \, \e{z-z_0} \, dz =
z_0 \left[ P\left( \frac{z_2-z_0}{\sigma} \right) -
P\left( \frac{z_1-z_0}{\sigma} \right) \right] & & \\
-\frac{\sigma}{\sqrt{2\pi}} \left\{ \e{z_2-z_0} - \e{z_1-z_0}
\right\}. & & \nonumber
\eea

\section{Scripts}

Numerical calculations, data analysis, and graphics were done with the
open-source, multi-platform statistical package
R version 2.12.0 (2010-10-15)~\cite{R}.  This documentation
was produced by integrating~\LaTeX\ with~R using the Sweave
package. The source code can be found in the file
\href{https://cdcvs.fnal.gov/redmine/documents/276}{dmcs.tar.gz}. Below
is the~R part of the code.

{\footnotesize
\begin{verbatim}
#line 71 "dmcs.Rnw"
Stangle('dmcs.Rnw', annotate=FALSE) # make dmcs.R for later inclusion

# collimator position vs. time, linear
Jc <- function(t, Jci=0.04, Jcf=0.05, dt=1){
  Jc <- 0*t + Jci # initial value
  subset <- dt<=t
  Jc[subset] <- Jcf # final value
  subset <- 0<t & t<dt
  Jc[subset] <- Jci + (Jcf-Jci)*t[subset]/dt # collimator moving
  return(Jc)
}

# slope of distribution vs. time, linear
Ac <- function(t, Ai=1, Af=0.9, dt=1){
  Ac <- 0*t + Ai # initial value
  subset <- dt<=t
  Ac[subset] <- Af # final value
  subset <- 0<t & t<dt
  Ac[subset] <- Ai + (Af-Ai)*t[subset]/dt # collimator moving
  return(Ac)
}

# initial, final, and asymptotic distributions, linear
flin <- function(J, A=1, J0=0.04){
  f <- 0*J # default outside range
  J0 <- abs(J0) # define edge as positive
  subset <- abs(J)<J0 # triangular with cusp at J=0
  f[subset] <- A*(J0 - abs(J[subset])) # linear
  return(f) }

# evolution of the distribution function (der=0)
# or its derivative (der=1), analytical formulas
fs <- function(J, t, Ai=1, Af=0.9, Jci=0.04, Jcf=0.05, D=1e-5, dt=1, der=0) {
  if(der!=0 & der!=1) stop("Derivative ('der') must be 0 or 1.")
  fs <- 0 # default outside collimators
  if(Jci>Jcf) IN<-TRUE else IN<-FALSE # inward or outward step?
  if(t<0){ # initial distribution
    if(der==0) fs <- flin(J, A=Ai, J0=Jci)
    if(der==1) fs <- -Ai
  } else{ # evolution
    Jco <- Jc(t, Jci=Jci, Jcf=Jcf, dt=dt) # collimator position
    Aco <- Ac(t, Ai=Ai, Af=Af, dt=dt) # slope of asymptotic solution
    if(0<=J & J<=Jco){ # inside collimators
      s <- sqrt(2*D*t) # Gaussian sigma
      # terms common to both inward and outward step
      if(der==0) fs.base <- -pnorm(-J/s) * (Ai*(Jci-J) - Aco*(Jco-J)) +
          pnorm((J-2*Jco)/s) * (Ai*(Jci+J-2*Jco) - Aco*(J-Jco)) +
          s/sqrt(2*pi) * (
            exp(-0.5*(J/s)^2) * (Aco-Ai) +
            (-1)*exp(-0.5*((2*Jco-J)/s)^2) * (Aco-Ai) )
      if(der==1) fs.base <- (Ai-Aco)*(pnorm(-J/s) + pnorm((J-2*Jco)/s)) +
          (Ai*Jci-Aco*Jco)/sqrt(2*pi)/s *
              (exp(-0.5*(J/s)^2)+exp(-0.5*((J-2*Jco)/s)^2))
      # add specific terms
      if(IN) { # inward step
        if(der==0) fs <- fs.base +
            (-1)*Ai*(Jci+J-2*Jco) +
            2 * pnorm((Jco-J)/s) * Ai * (Jci-Jco)
        if(der==1) fs <- fs.base +
            (-1)*Ai + 2*Ai*(Jco-Jci)*exp(-0.5*((Jco-J)/s)^2)/sqrt(2*pi)/s
      } else { # outward step
        if(der==0) fs <- fs.base +
          pnorm((Jci-J)/s) * Ai * (Jci-J) +
          (-1)*pnorm((Jci+J-2*Jco)/s) * Ai * (Jci+J-2*Jco) +
          s/sqrt(2*pi) * (
            exp(-0.5*((Jci-J)/s)^2) * Ai +
            (-1)*exp(-0.5*((Jci+J-2*Jco)/s)^2) * Ai )
        if(der==1) fs <- fs.base +
            (-1)*Ai*(pnorm((Jci-J)/s)+pnorm((Jci+J-2*Jco)/s))
      }
    }
  }
  return(fs)
}
fv <- Vectorize(fs, vectorize.args=c("J","t")) # vectorize the function

# loss rate vs. time
# particle flux at the collimator L = -D * (df/dJ)|J=Jc
# with conversion constant k and background B
loss.rate <- function(t, Ai=1, Af=0.9, Jci=0.04, Jcf=0.05, D=1e-5, dt=1,
                      k=1, B=0) {
  L <- -Ai + 0*t # default before step
  subset <- t>=0 # evolution
  s <- sqrt(2*D*t[subset]) # Gaussian sigmas
  Jco <- Jc(t[subset], Jci=Jci, Jcf=Jcf, dt=dt) # collimator positions
  Aco <- Ac(t[subset], Ai=Ai, Af=Af, dt=dt) # slopes
  if(Jci>Jcf) IN<-TRUE else IN<-FALSE # inward or outward?
  # common terms
  L[subset] <- 2*(Ai-Aco)*pnorm(-Jco/s) +
    2/sqrt(2*pi)/s * (Ai*Jci-Aco*Jco) * exp(-0.5*(Jco/s)^2)
  if(IN) {
      L[subset] <- L[subset] - Ai - 2/sqrt(2*pi)/s * Ai * (Jci-Jco)
  } else {
      L[subset] <- L[subset] - 2 * Ai * pnorm((Jci-Jco)/s) }
  return(-k*D*L+B)
}

# sample data
# read collimator position and loss rate data
ReadACNETData <- function(fn="CFBI_NG.txt.gz") {
cat("Reading file ", fn, "...\n")
dn <- gsub(".txt.gz", "", fn) # name of data frame
t0 <- scan(fn, what="character", skip=1, nlines=1) # get initial time
t0 <- t0[6]
h <- as.numeric(substr(t0,1,2))
m <- as.numeric(substr(t0,4,5))
s <- as.numeric(substr(t0,7,8))
assign("T0", h+m/60+s/3600, envir=.GlobalEnv)
cat("Start time: ", T0, " h\n")
cnam <- scan(fn, what="character", skip=2, nlines=1, sep="\t") # column names
cnam <- substr(cnam, 3, 12) # strip logger names
cnam <- gsub("[^[:alnum:]]", "", cnam) # remove non alphanumeric characters
if(length(cnam)<72) cnam <- gsub("0", "", cnam) # remove 0s from scalars
cnam[seq(1,length(cnam),by=2)] <-
  paste("t", cnam[seq(2,length(cnam),by=2)], sep="") # times
assign(dn, read.table(fn, header=FALSE, skip=3, colClasses="numeric",
                      na.strings=c("NA","999999"), col.names=cnam),
       envir= .GlobalEnv) # fill data frame
}
ReadACNETData('Store8749_CP.txt.gz')
ReadACNETData('Store8749_TLF480_15Hz.txt.gz')

tc <- Store8749_CP$tF48VCP/3600 + T0 # time stamps of coll. positions [h]
cpmil <- Store8749_CP$F48VCP # collimator position [mils]
cp0 <- 366 # collimator offset (arbitrary)
cp <- (cpmil - cp0) * 25.4 # collimator position [um]

tT1 <- Store8749_TLF480_15Hz$tLF48/3600 + T0 # time stamps of losses [h]
T1 <- Store8749_TLF480_15Hz$LF48 # loss monitor [Hz]

# plot setup
options(SweaveHooks=list(fig=function() { # common plot options
    pt<-10; fo<-"Times"
    ps.options(pointsize=pt, family=fo)
    pdf.options(pointsize=pt, family=fo)
    par(oma=rep(0,4), mar=c(4.2,4.2,0.2,0.2)) }))


#line 217 "dmcs.Rnw"
ti <- 9.105 # initial time [h]
tf <- 9.294 # final time
plot.unit <- 60
xl <- c(0, tf-ti)*plot.unit
par(mar=c(0, 3.2, 0, 0), oma=c(3.2, 0, 0.2, 0.2), mgp=c(2,1,0), tcl=-0.25)
layout(matrix(1:2, nrow=2, ncol=1))
x <- subset((tT1-ti)*plot.unit, ti<=tT1 & tT1<=tf) # time stamps and loss rates
y <- subset( T1, ti<=tT1 & tT1<=tf)
ii <- round(seq(1, length(x), length=2048)) # choose a few representative points
x <- x[ii]; y <- y[ii]
plot(x, y, type="l", xlim=xl, axes=FALSE, xlab="",
     ylab="Loss rate [arb. units]")
axis(2); box()
x <- subset((tc-ti)*plot.unit, ti<=tc & tc<=tf) # collimator positions
y <- subset(cp, ti<=tc & tc<=tf)
plot(x, y, type="o", pch=20, xlim=xl, xlab="",
     ylab=expression(paste("Collimator position [", mu*m, "]")))
mtext("Time [min]", side=1, line=2, outer=TRUE)


#line 365 "dmcs.Rnw"
# sample parameters
Jci <- 0.040; Jcf <- 0.050 # initial and final collimator actions [um]
dt <- 1 # time to move collimator from Jci to Jcf [s]
Ai <- 1 # initial slope of distribution function [um^-2]
Af <- Ai*(Jci/Jcf) # final slope
tt <- seq(0, dt, length=5) # choose times
JJ <- Jc(tt, Jci=Jci, Jcf=Jcf, dt=dt) # calculate coll. pos.
AA <- Ac(tt, Ai=Ai, Af=Af, dt=dt) # calculate slopes
co <- gray(seq(0.6, 0, length=length(tt))) # define colors, grayscale
Jmin <- 0; Jmax <- max(Jci,Jcf) # plot range
plot(c(Jmin,Jmax),
     c(0, max(Jci*Ai,Jcf*Af)),
     type="n",
     xlab=expression(paste("Action [", mu, "m]", sep="")),
     ylab=expression(paste("Distribution function [", mu*m^{-1}, "]", sep="")))
abline(v=JJ, col=co, lwd=2) # plot collimator positions
for(i in seq(along=tt)){ # plot asymtotic functions
  curve(flin(x, A=AA[i], J0=JJ[i]), Jmin, Jmax, add=TRUE, col=co[i], lwd=2) }
legend("bottomleft", bty="n", # describe times for each curve
       title="Time [s]", title.col="black",
       legend=format(tt, di=2), col=co, text.col=co)
xx <- 0.5*(Jmin+Jmax) # labels curves
text(xx, flin(xx, A=Ai, J0=Jci), expression(f[0](J)), pos=2)
text(xx, flin(xx, A=Af, J0=Jcf), expression(f[infinity](J)), pos=4)
xx <- 0.6*Jmax; yy <- 0.1*max(Jci*Ai, Jcf*Af)
text(xx, yy, expression(f[a](J,t)), pos=2)
xx2 <- 0.7*Jmax
lines(x=c(xx, xx2), y=c(yy, flin(xx2, A=AA[2], J0=JJ[2])))
caption.text <- "no" # text for LaTeX caption
if(Jcf<Jci) caption.text <- "inward"
if(Jcf>Jci) caption.text <- "outward"


#line 554 "dmcs.Rnw"
# sample parameters
J1 <- 0.040; J2 <- 0.050 # initial and final collimator actions [um]
dt <- 1 # time to move collimator from Jci to Jcf [s]
A1 <- 1 # initial slope of distribution function [um^-2]
A2 <- A1*(J1/J2) # final slope
D <- 1e-5 # diffusion coefficient [um^2/s]
# choose times to plot
tscale.short <- abs(J1-J2)^2/D
tscale.long <- min(J1,J2)^2/D
tt <- c(seq(0, dt, length=5), # collimator moving
        c(3*dt, tscale.short, 2*tscale.short), # short term after
        c(5*tscale.short, tscale.long, 2*tscale.long)) # long term after
co <- gray(seq(0.6, 0, length=length(tt))) # define colors, grayscale
Jmin <- 0; Jmax <- max(J1,J2) # plot range
layout(matrix(1:2, nrow=1, ncol=2)) # plot layout
for(i in 1:2) {
  if(i==1){ # left plot, inward
    Jci<-J2; Jcf<-J1; Ai<-A2; Af<-A1}
  if(i==2){ # right plot, outward
    Jci<-J1; Jcf<-J2; Ai<-A1; Af<-A2}
  JJ <- Jc(tt, Jci=Jci, Jcf=Jcf, dt=dt) # calculate coll. pos.
  AA <- Ac(tt, Ai=Ai, Af=Af, dt=dt) # calculate slopes
  ylabel <- expression(paste("Distribution function [", mu*m^{-1}, "]", sep=""))
  if(i==2) ylabel <- ""
  plot(c(Jmin,Jmax),
     c(0, max(Jci*Ai,Jcf*Af)),
     type="n",
     xlab=expression(paste("Action [", mu, "m]", sep="")),
     ylab=ylabel)
abline(v=JJ, col=co, lwd=2) # plot collimator positions
for(i in seq(along=tt)){ # plot asymptotic functions
  curve(fv(J=x, t=tt[i], Ai=Ai, Af=Af, Jci=Jci, Jcf=Jcf, D=D, dt=dt),
        Jmin, Jmax, add=TRUE, col=co[i], lwd=2) }
legend("bottomleft", bty="n", # describe times for each curve
       title="Time [s]", title.col="black",
       legend=formatC(tt, drop0=TRUE), col=co, text.col=co, cex=0.8)
}


#line 668 "dmcs.Rnw"
# sample parameters
J1 <- 0.040; J2 <- 0.050 # initial and final collimator actions [um]
dt <- 1 # time to move collimator from Jci to Jcf [s]
D <- c(1e-6, 1e-5, 1e-4) # show the effect of different diffusion coeff.
k <- 1; B <- 0
tscale.short <- abs(J1-J2)^2/min(D) # estimate time scales
tscale.long <- min(J1,J2)^2/min(D)
tmax <- tscale.short/2
co <- gray(seq(0.5, 0, length=length(D))) # define colors, grayscale
Jmin <- 0; Jmax <- max(J1,J2) # plot range
layout(matrix(1:2, nrow=1, ncol=2)) # plot layout
for(i in 1:2) {
  if(i==1){ # left plot, inward
    Jci<-J2; Jcf<-J1
    ylabel <- "Loss rate [arb. units]"
    legend.pos <- "topright"}
  if(i==2){ # right plot, outward
    Jci<-J1; Jcf<-J2
    ylabel <- ""
    legend.pos <- "bottomright"}
  ADD <- FALSE
  for(j in 1:length(D)){
    Ai <- 1/D[j] # scale slope to keep steady-state rates constant
    Af <- Ai*Jci/Jcf
    if(j>1) ADD <- TRUE
    curve(loss.rate(t=x, Jci=Jci, Jcf=Jcf, Ai=Ai, Af=Af, D=D[j],
                    dt=dt, k=k, B=B), -0.1*tmax, tmax,
          xlab="Time [s]", ylab=ylabel, add=ADD, lwd=2, col=co[j])}
  if(i==1) legend(legend.pos, bty="n", # print diffusion rate for each curve
     title=expression(paste("Diffusion rate [", mu*m^2/s, "]")),
     title.col="black",
     legend=formatC(D, format="E", di=1), col=co, text.col=co)
}


\end{verbatim}
}

\end{document}